\documentclass[final,3p,times]{elsarticle}

 \usepackage{graphicx}

\usepackage{amssymb}

\journal{Comptes Rendus de l'Acad{\'e}mie des Sciences (Comptes Rendus Physique)}

\begin{document}

\begin{frontmatter}



\title{Probing the order parameter symmetry in the cuprate high temperature superconductors by SQUID microscopy}


\author{John R. Kirtley}

\address{Center for Probing the Nanoscale, Stanford University, Stanford, CA, 93404 USA}
\ead{jkirtley@stanford.edu}
\ead[http://www.kirtleyscientific.com]{kirtleyscientific.com}

\begin{abstract}
The orbital component of the order parameter in the cuprate high-T$_c$ cuprate superconductors is now well established, in large part because of phase sensitive tests.  Although it would be desirable to use such tests on other unconventional superconductors, there are a number of favorable factors associated with the properties of the cuprates, and a number of technical advances, that were required for these tests to be successful. In this review I will describe the development of phase sensitive pairing symmetry tests using SQUID microscopy, underlining the factors favoring these experiments in the cuprates and the technical advances that had to be made. 

La composante orbitale du param{\'e}tre d'ordre dans les cuprates {\'a} haute T$_c$ supraconducteurs cuprate est maintenant bien {\'e}tabli, en grande partie ˆ cause  des tests sensibles  {\'a} la phase. Bien qu'il serait souhaitable d'utiliser de tels tests sur d'autres supraconducteurs non conventionnels,  les propri{\'e}t{\'e}s des cuprates etaient favorables, et en plus un certain nombre de progr{\'e}s techniques, ont {\'e}t{\'e} n{\'e}cessaires pour transformer ces essais en reussite. Dans cette article de revue, je vais vous d{\'e}crire le d{\'e}veloppement des tests de phase permettant d'identifier  la sym{\'e}trie du parametre d'ordre en utilisant la microscopie a SQUID, en soulignant les facteurs favorisant ces exp{\'e}riences dans les cuprates et les progr{\'e}s techniques qui ont d{\^u} {\^e}tre apport{\'e}es.
\end{abstract}


\end{frontmatter}

It is difficult to convey the excitement generated by the discovery almost 25 years ago of superconductivity at elevated temperatures in the perovskite cuprates \cite{bednorz1986pht,wu1987san}. This discovery stimulated an intense debate about the mechanism for high-temperature superconductivity \cite{anderson1987rvb,hirsch1988pit,emery1988mht,schrieffer1988sbm,varma1989pns,monthoux1992wct,newns1995vhs,scalapino1998sce}. Much of this debate concerned the symmetry of the orbital component of the Cooper pairing order parameter, which was considered by some to be key to understanding the pairing mechanism. Although it was proposed quite early that the cuprates might have unconventional Cooper pairing symmetry \cite{scalapino1986dwp,lee1987wit,gros1987sil,monthoux1992wct}, there was strong resistance to this idea. This strong resistance is, in retrospect, somewhat surprising given that unusual pairing order parameters had been discussed for a number of years in connection with the heavy Fermion superconductors \cite{sigrist1991ptu}. Gradually evidence for a primarily $d_{x^2-y^2}$ pairing symmetry emerged from a number of non-phase sensitive experimental techniques \cite{scalapino1995cdp}. However, the pairing symmetry in the cuprates was still a very controversial topic before the advent of phase sensitive pairing symmetry experiments.

\begin{figure}
        \begin{center}
                \includegraphics[width=10cm]{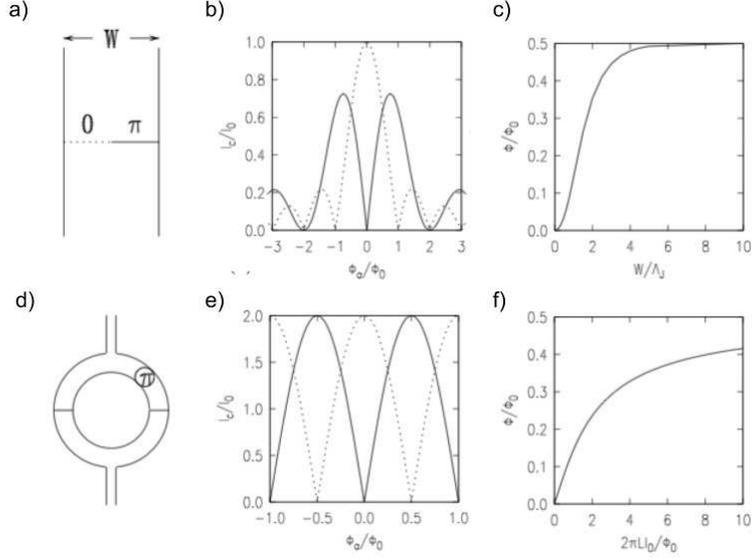}
        \end{center}
        \caption{Two classes of phase sensitive pairing symmetry tests. (a) 0-$\pi$ junction. (b) Critical current for a symmetric 0-$\pi$ junction  (solid line) and a conventional ($0$) junction (dashed line) in the short junction limit.  (c) The spontaneously generated magnetic flux in a symmetric 0-$\pi$ junction vs. $W/\lambda_J$. (d)  A two-junction SQUID with an intrinsic $\pi$ phase shift. (e) The critical current of a symmetric, two-junction $\pi$-SQUID  (solid line) and $0$-SQUID (dashed line) in the limit $2\pi L I_0/\Phi_0<<1$. (f) The spontaneously generated flux in a symmetric, two-junction $\pi$-SQUID as a function of  $2\pi L I_0/\Phi_0$.}
        \label{fig:pitheoryh}
\end{figure}

The theoretical basis for pairing symmetry experiments is well established \cite{bulaevskii1977ssw,geshkenbein1987vhm,sigrist1992peh}. They are performed on superconducting junctions or SQUIDs (Figure \ref{fig:pitheoryh}(a,d)) with intrinsic phase shifts intentionally introduced  through the momentum dependence of the Cooper pairing wavefunction. These phase shifts are inferred either through the magnetic field dependence of the critical current \cite{wollman1993eds,wollman1995edp,vanHarlingen1995pst} (Figure \ref{fig:pitheoryh}(b,e)), or by detecting a spontaneously generated magnetic flux \cite{tsuei1994psf,tsuei2000psc} (Figure \ref{fig:pitheoryh}(c,f)), in the resulting devices. Pairing symmetry tests take advantage of the fact that the Josephson supercurrent across a junction is set by the amplitudes and phases of the normal components of the pairing order parameter on the two sides of the junction \cite{sigrist1992peh}. A junction which has a shift of $\pi$ in the intrinsic phase drop across the junction between two of its sections is called a $0-\pi$ junction. A ring of superconducting material with one or more Josephson weak links with an intrinsic integrated phase change of $\pi$ around the ring is called a $\pi$-ring. The critical current of a conventional ($0$) Josephson junction and a $0-\pi$ junction with equal lengths in the short-junction limit $W/\lambda_J<<1$, where $W$ is the width of the junction, $\lambda_J=\sqrt{\hbar/2e\mu_0dj_c}$ is the Josephson penetration depth , $d$ is the spacing between superconducting faces making up the junction, and $j_c$ is the Josephson critical current per area of the junction, is given by (Figure \ref{fig:pitheoryh}(b)):
\begin{equation}
I_c(\Phi) = \left \{
\begin{array}{ll}
I_0 | \sin(\pi\Phi/\Phi_0)/(\pi\Phi/\Phi_0)| & {\rm 0 \,\, junction} \nonumber \\
I_0 | \sin^2(\pi\Phi/2\Phi_0)/(\pi\Phi/2\Phi_0)| & {\rm 0-\pi \,\, junction},
\end{array} \right.
\label{eq:junction_ic}
\end{equation}
where $I_0 = j_c t W$, $t$ is the thickness of the junction, $\Phi$ is the magnetic flux threading the junction, and $\Phi_0=h/2e$ is the superconducting flux quantum. Similarly the critical current of a symmetric two-junction SQUID, in the limit $2\pi LI_0/\Phi_0 << 1$, where $L$ is the inductance of the SQUID loop and $I_0$ is a junction's critical current, is given by (Figure \ref{fig:pitheoryh}(e)). 
\begin{equation}
I_c(\Phi) = \left \{
\begin{array}{ll}
2 I_0 | \cos(\pi\Phi/\Phi_0)| & {\rm 0-SQUID} \nonumber \\
2 I_0 | \sin(\pi\Phi/\Phi_0)| & {\rm \pi-SQUID},
\end{array} \right.
\label{eq:squid_ic}
\end{equation}
In the alternate limits of long junctions ($W/\lambda_J >>1$) or large $LI_0$ products ($2\pi LI_0/\Phi_0 >> 1$) the modulation amplitude of the device critical current with magnetic field approaches zero, but spontaneous circulating supercurrents generate magnetic flux approaching $\Phi_0/2$ for both $0-\pi$ junctions and $\pi$ rings as $W/\lambda_J$ or $2\pi L I_0/\Phi_0 \rightarrow \infty$ (Figure \ref{fig:pitheoryh}(c,f)). It is these spontaneous fluxes that are most often measured for SQUID microscope based pairing symmetry tests. In addition, in a $\pi$-ring at zero externally applied magnetic field, with sufficiently high $LI_0$ products, there is a ladder of allowed flux states, spaced by $\Phi_0$, with the lowest energy state being two-fold degenerate and having $\pm \Phi_0/2$ flux. A conventional $0$-ring under the same conditions has a ladder of allowed flux states spaced by $\Phi_0$ but centered at zero flux.

The first experimental pairing symmetry tests on the cuprates were measurements of the magnetic field dependence of the critical current of corner and edge junctions and SQUIDs fabricated between YBa$_2$Cu$_3$O$_{7-\delta}$ (YBCO) single crystals and Pb counterelectrodes \cite{wollman1993eds,wollman1995edp,vanHarlingen1995pst}. These experiments were soon followed by similar experiments with point contact YBCO-Nb SQUIDs by Brawner and Ott \cite{brawner1994eus}. These experiments depended on three favorable factors for their success. The first was the availability of high quality single crystals of YBCO. The second was that both the Fermi surface (nearly cylindrical) and the pairing symmetry (predominantly $d_{x^2-y^2}$) were particularly simple in the cuprates, especially when compared with what has been proposed for Sr$_2$RuO$_4$ \cite{mackenzie2003ssp}, the heavy Fermion \cite{sigrist1991ptu}, or the pnictide \cite{mazin2009psp} superconductors. Although some of the early pairing symmetry experiments were on untwinned single crystals \cite{vanHarlingen1995pst}, a third favorable factor  was that the order parameter has odd reflection symmetry in the twin boundaries in the cuprates \cite{walker1996jth}. This meant that the $d_{x^2-y^2}$ component of the order parameter maintained the same sign in the same direction across twin boundaries, making pairing symmetry experiments possible on twinned samples.

While the magnetic interferometry experiments ultimately proved to be correct in their qualitative conclusion that the cuprates had predominantly $d_{x^2-y^2}$ pairing symmetry, they were initially met with skepticism in some quarters \cite{klemm1994coe,klemm1998wsh}. There was at least some rational basis for this skepticism: This class of experiments depends on conservation of Cooper pair momentum across the tunnel barrier. Since the original experimental junctions were formed on macroscopic faces of a single crystal, it was not immediately obvious that this would be the case. Flux trapping in annular Josephson junctions between $s$-wave superconductors was known to produce interference patterns qualitatively similar to those observed in the magnetic interferometry pairing symmetry experiments \cite{cristiano1999mpa}. The observed interference patterns were far from ideal, self-field effects had to be corrected for, and there was a large uncertainty in the phase shift inferred between adjacent faces of the YBCO single crystals \cite{wollman1993eds,brawner1994eus}. Sample temperatures could only be varied over a limited range, since the junctions and SQUIDs were made with Pb or Nb as counter-electrodes. Many of these difficulties were ultimately eliminated in later magnetic interferometry experiments using YBCO tricrystal samples \cite{schulz2000dra}.

The SQUID microscope tests I will describe in this review avoid many of the difficulties originally encountered in the magnetic interference tests. The biggest advantage of SQUID microscope experiments is that all of the magnetic fields in the sample region are sensitively and quantitatively imaged, eliminating the possibility that the observed effects are due to flux trapping. Regions (or rings) that are expected to spontaneously generate flux can be directly compared with those that are not. Sample geometries can be varied, in the case of the ramp-edge junctions nearly continuously. Sample compositions, dopings, and temperatures can be varied over a broad range. 

\begin{figure}[tb]
        \begin{center}
                \includegraphics[width=12cm]{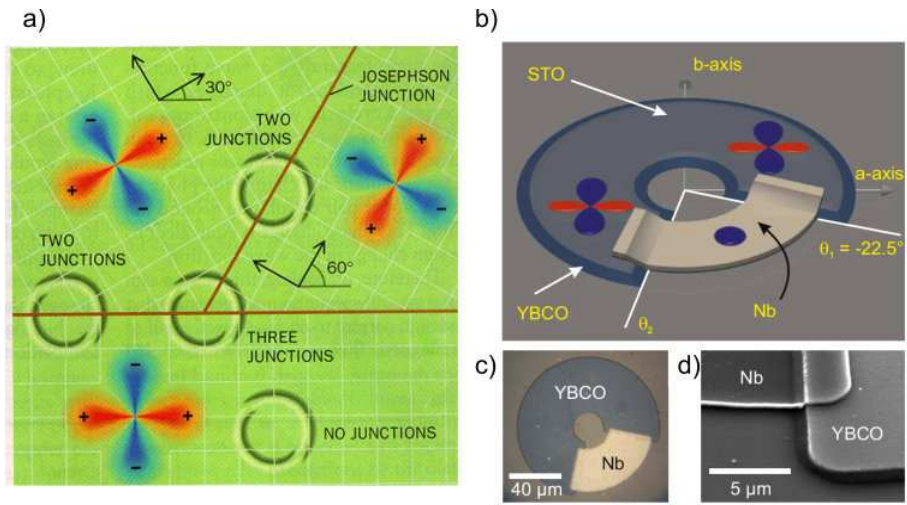}
        \end{center}
       \caption{Two geometries for SQUID microscope tests of pairing symmetry. a) Tricrystal geometry \cite{tsuei1994psf}.  b) Ramp-edge junction geometry \cite{kirtley2006arp}. c) Optical micrograph of a completed YBCO-Nb ring with ramp-edge junctions. d) Scanning electron micrograph of a YBCO-Nb ramp-edge junction.}
        \label{fig:geometries}
\end{figure}

Two of the sample geometries used for pairing symmetry experiments with SQUID microscopy are shown in Figure \ref{fig:geometries}. In the first (Figure \ref{fig:geometries}(a)), thin film cuprate superconductors were epitaxially grown on a tricrystal substrate of SrTiO$_3$ with a geometry chosen to produce an odd number of sign changes of the normal component of a $d_{x^2-y^2}$ order parameter for a ring circling the central point \cite{tsuei1994psf}. There were several favorable factors that made these experiments possible. First, the cuprate superconductors grow epitaxially on SrTiO$_3$, and grain boundaries between crystallites of different orientation can be made simply by growing on bi-, tri-, and quad-crystals with the appropriate geometry \cite{hilgenkamp1992gbh}. Second, grain boundaries in the cuprate superconductors are Josephson weak links at sufficiently high misorientation angles \cite{hilgenkamp1992gbh}. Third, as mentioned above, although thin films of the cuprates are highly twinned, the order parameter in the cuprates has odd-reflection symmetry across twin boundaries. Fourth, Josephson tunneling across grain boundaries apparently favors Cooper pairs with low momentum parallel to the grain boundary interface, so that the tricrystal geometry produces $\pi$-rings as designed for the cuprate superconductors.

\begin{figure}[tb]
        \begin{center}
                \includegraphics[width=12cm]{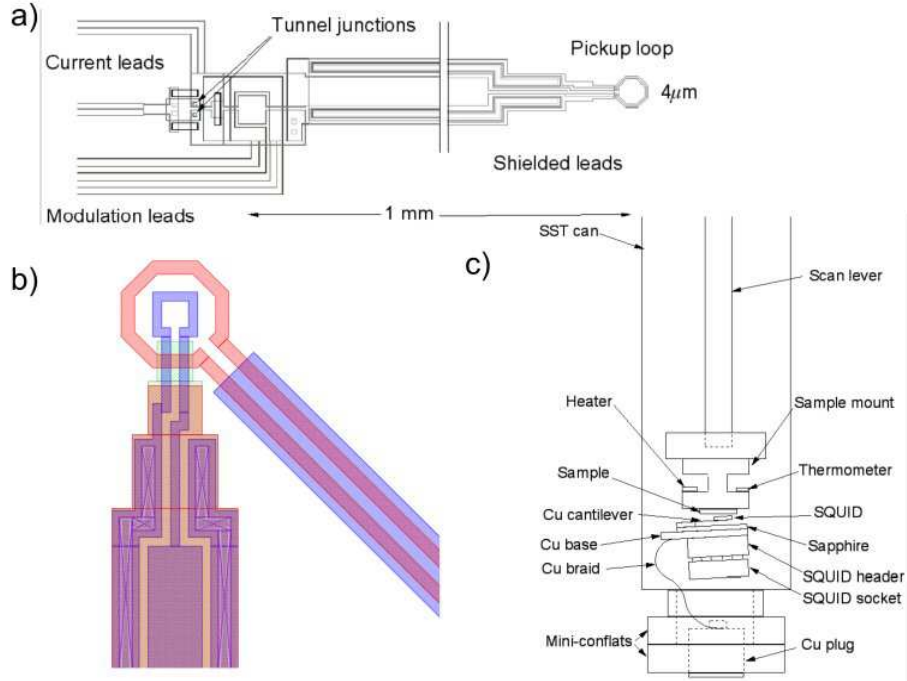}
        \end{center}
       \caption{(a) Schematic layout for a scanning SQUID sensor with integrated 4$\mu$m diameter pickup loop. (b) Schematic layout of the pickup loop area of a scanning SQUID susceptometer. (c) Schematic diagram of a variable sample temperature SQUID microscope.}
        \label{fig:squid_schematics}
\end{figure}

The second geometry for SQUID microscopy pairing symmetry tests (Figure \ref{fig:geometries}(b)) used ramp-edge junctions formed between YBCO and Nb arcs to form two-junction rings. One of the junction normal axes was held fixed at $\theta_1=-22.5^o$ relative to the YBCO $a$-axis, while the other junction angle $\theta_2$ was varied from ring to ring. Pairing symmetry test experiments using this geometry depended on two important technical advances. First, high quality junctions between YBCO and Nb can be grown if the YBCO is etched back after patterning, regrown in situ, and the junctions are completed with layers of Au and Nb, also in situ \cite{smilde2002etr}. Second, YBCO thin films can be grown untwinned on a slightly miscut (vicinal) SrTiO$_3$ substrate \cite{dekkers2003myb}. This allowed measurements of the anisotropy of the YBCO order parameter between the $a$ and $b$ in-plane crystalline directions.

It was recognized almost as soon as the first demonstration of the Josephson effect \cite{anderson1963poj} and superconducting quantum interference effects \cite{jaklevic1964qie} that magnetic fields could be imaged by scanning SQUIDs relative to samples \cite{zimmerman1964qfp}, and superconducting vortices were imaged using SQUID microscopy as early as the 1980's \cite{rogers1983deo}. However, high spatial resolution SQUID microscopes were just being developed \cite{mathai1993odm,black1993mmu,vu1993imv,vu1993dis,ma1993hri,kirtley1995hrs} when the pairing symmetry debate was at its peak. There are two strategies for doing high spatial resolution SQUID microscopy. The first is to make small SQUIDs \cite{wernsdorfer1997een,veauvy2002smus,troeman2007nbn,granata2008ism,hao2008man,finkler2010san}. I will concentrate in this review on the second, to make more conventional (large) SQUIDs that have a small, integrated, and well shielded pickup loop \cite{vu1993imv,vu1993dis,ketchen1995dap,kirtley1995hrs}. Small SQUIDs have the advantage of simplicity, while integrated pickup loop SQUIDs have the advantage of ease of flux modulation.

Schematics of SQUID sensors and microscopes used for pairing symmetry tests are shown in Figure \ref{fig:squid_schematics}. Figure \ref{fig:squid_schematics}(a) shows a typical layout for a scanning SQUID magnetometer sensor. These sensors use low-T$_c$ Nb-Al$_2$O$_3$-Nb trilayer junctions \cite{gurvitch1983hqf}
 with several Nb ground and wiring levels, using optical lithography with down to 0.7$\mu$m wire widths and spacings. The junctions, resistive shunts, and modulation coil are well separated from the pickup loop, which is integrated into the SQUID through a thin-film coaxial sheath, so that only the pickup loop is sensitive to the sample magnetic fields. Recently pickup loop sizes as small as 0.6$\mu$m in diameter have been defined using focussed ion beam lithography \cite{koshnick2008ats}. Figure \ref{fig:squid_schematics}(b) shows the layout for the pickup loop area of a scanning SQUID susceptometer \cite{gardner2001ssq,huber2008gms}. In these sensors a one-turn field coil is integrated into the pickup loop area. The SQUIDs are fabricated on silicon substrates, which have a corner close to the pickup loop formed by mechanical polishing, sawing, or chemical etching, and mounted on a flexible cantilever so that the SQUID substrate is nearly parallel to the sample surface, with the pickup loop as close as possible to the sample. Scanning is done either using mechanical motors or piezoelectric scanners. Figure \ref{fig:squid_schematics}(c) illustrates the design of a variable sample temperature SQUID microscope \cite{kirtley1999vst}, in which the SQUID is closely thermally anchored to a cold bath, so that it will remain superconducting while the sample is warmed. In this microscope the scanning was done with a lever mechanism, with the sample attached to the bottom of the lever, and mechanical motors operating the other end of the lever at room temperature at the top of the dewar.

\begin{figure}[tb]
        \begin{center}
                \includegraphics[width=12cm]{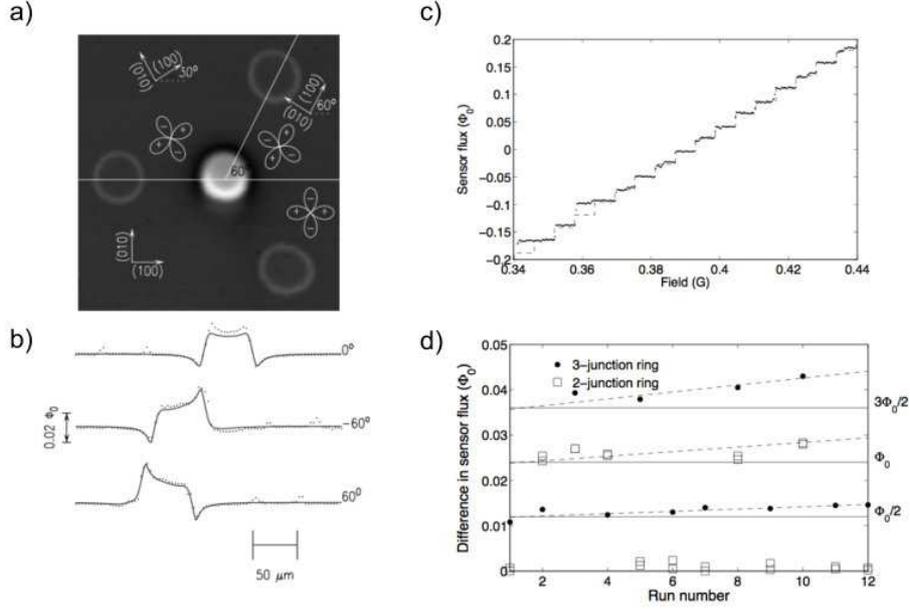}
        \end{center}
        \caption{SQUID microscope results from YBCO tricrystal experiments \cite{tsuei1994psf}. a) SQUID microscopy image of a YBCO tricrystal ring sample.  b) Cross-sections (dots) through the central ring at angles relative to the horizontal as indicated.  The solid lines are calculations assuming the central ring has $\Phi_0/2$ total flux.  c) Flux through the SQUID sensor, centered above the 3-junction ring, as a function of externally applied field perpendicular to the sample plane. d) Plot of the difference in flux through the SQUID sensor centered above the 3-junction ring (dots) and the 2-junction rings (squares) minus that above the ring with no junctions, for 12 cooldowns.}
        \label{fig:triprl}
\end{figure}

Experimental results from the original tricrystal ring scanning SQUID microscope experiments \cite{tsuei1994psf} are shown in Figure \ref{fig:triprl}. For Fig. \ref{fig:triprl}(a) the sample was cooled in zero field and imaged at 4.2K using a SQUID with a 10$\mu$m diameter pickup loop. The grain boundaries are indicated by solid lines overlaid on the magnetic image. The two 2-junction rings and the ring with no junctions have no spontaneous circulating current, and are visible through mutual inductance between the ring wall and the SQUID pickup loop. The central, 3-junction ring has spontaneous flux generated in it. It is immediately clear from this image that the central ring has more magnetic flux in it than the control rings. However, it was necessary to determine how much. An advantage to SQUID microscopy is that it is easy to do absolute calibrations of the magnetic field and flux threading the pickup loop, because the critical current of the SQUID is periodic in the flux threading it, with period $\Phi_0=h/2e$, and the sensor geometry is well known. Fig. \ref{fig:triprl}(b-d) shows three methods for determining that there is $\Phi_0/2$ of total flux through the 3-junction ring when it is cooled in zero field. The dots in Fig. \ref{fig:triprl}(b) are cross-sections through the magnetic image of Fig. \ref{fig:triprl}(a) at angles relative to the horizontal as indicated. The solid lines are calculations, using the known geometry of the ring and pickup loop, and assuming the ring has $\Phi_0/2$ total flux in it. Fig. \ref{fig:triprl}(c) shows SQUID flux vs applied field characteristics when the SQUID pickup loop is centered on the 3-junction ring. Magnetic flux enters the ring in quantum steps, resulting in steps in the SQUID sensor output. Each step corresponds to $\Phi_0$ of flux in the ring. The dashed line is made up of evenly spaced steps in both field and flux.  Fig. \ref{fig:triprl}(d) shows the amplitude of the SQUID sensor flux signal when the pickup loop is centered on the 3-junction and 2-junction rings, minus that above the 0-junction ring, for 12 separate cooldowns in slightly different fields. The 2-junction ring fluxes are close to integer multiples of an amplitude corresponding to $\Phi_0$, while the 3-junction ring fluxes are half-integer multiples of this value. The linear increase in signal visible from run to run was attributed to wear in the SQUID substrate, which was in direct contact with the sample.

\begin{figure}[tb]
        \begin{center}
                \includegraphics[width=12cm]{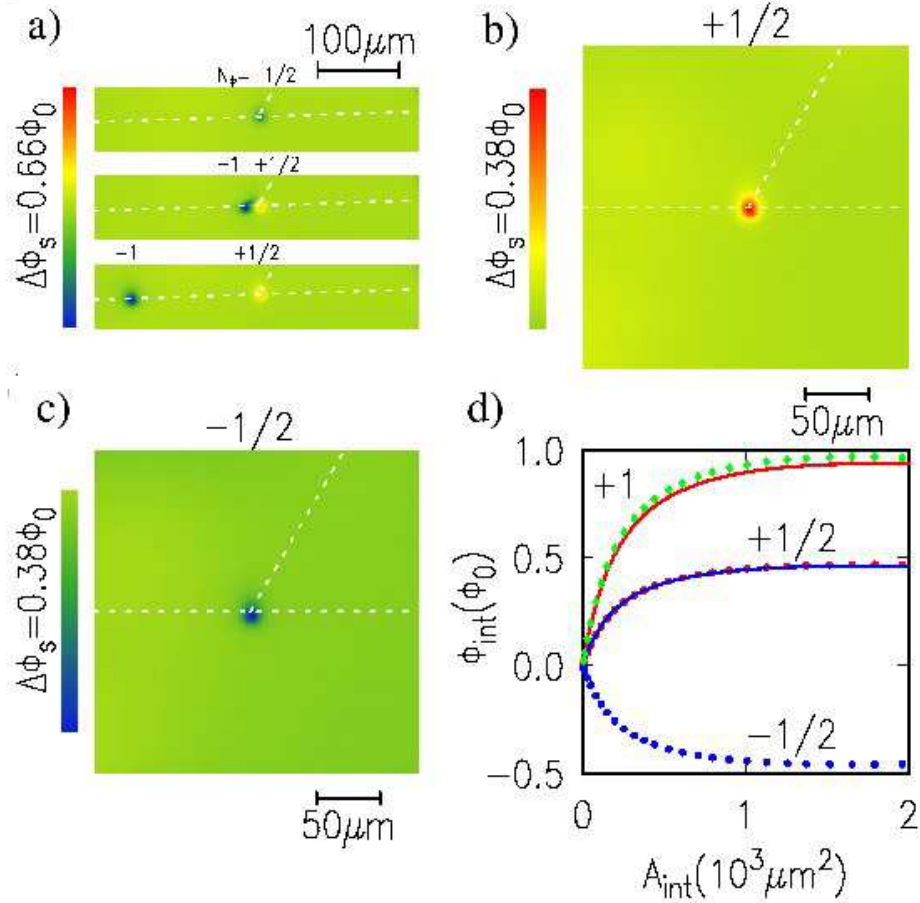}
        \end{center}
        \caption{SQUID microscopy images of the central region of a YBCO tricrystal sample. (a) illustrates the inversion of a N=-1/2 vortex using a locally applied field pulse. (b) and (c) are a N=+1/2 and a N=-1/2 Josephson vortex at the tricrystal point respectively. (d) shows the integrated flux as a function of integration area of a circle centered on the peak flux position for N=$\pm$1/2 and N=1 vortices.}
        \label{fig:dopesci1p}
\end{figure}

An advantage of the tricrystal technique is that it allows different geometries to be tested, simply by using different SrTiO$_3$ tricrystal substrates. In this way a symmetry independent mechanism for the half-flux quantum effect in the tricrystal samples \cite{kirtley1995sop}, and the simplest version of extended $s$-wave pairing symmetry \cite{tsuei2000psc}, were eliminated. A further simplification resulted when it was realized that it was not necessary to pattern the cuprate films into rings. The tricrystal point forms a half-flux quantum Josephson vortex when coated with a thin film of a $d_{x^2-y^2}$ superconductor \cite{kirtley1996dii}. This half-flux quantum vortex is easy to distinguish from integer vortices, and is present when the samples are cooled in zero field. This facilitated the demonstration of  predominantly $d_{x^2-y^2}$ pairing symmetry in  {Bi}$_2${Sr}$_2${CaCu}$_2${O}$_{8+\delta}$ \cite{kirtley1996hif}, Tl$_2$Ba$_2$CuO$_{6+d}$ \cite{tsuei1996hiq}, La$_{1.85}$Sr$_{0.15}$CuO$_y$ \cite{tsuei2004rdp}, the electron-doped cuprate superconductors Nd$_{1.85}$Ce$_{0.15}$CuO$_{4-y}$  and Pr$_{1.85}$Ce$_{0.15}$CuO$_{4-y}$ \cite{tsuei2000pse}, and hole-doped cuprates over a broad doping range \cite{tsuei2004rdp}. Predominantly $d_{x^2-y^2}$ pairing symmetry in YBCO was confirmed using scanning SQUID magnetometry by the Wellstood group at the University of Maryland \cite{mathai1995ept}, the Iguchi group at the Tokyo Institute of Technology \cite{Sugimoto2002tdh}, and the Lombardi group at Chalmers University \cite{cedergren2010ibs}.

An advantage of doing pairing symmetry tests with a scanning SQUID susceptometer is that the half-flux quantum vortex, which is doubly degenerate, can be manipulated by applying local magnetic fields \cite{tsuei2004rdp}. An example is shown in Figure \ref{fig:dopesci1p}. In Figure \ref{fig:dopesci1p}(a) a N=-1/2 Josephson vortex at the tricrystal point (upper panel) is inverted by passing a 5 mA pulse of current through a susceptometer field coil to form a +1/2 vortex, creating also a N=-1 Josephson vortex in the horizontal grain boundary (middle image). The N=-1 Josephson vortex is dragged from the tricrystal point by moving the sensor parallel to the grain boundary while applying a current of 4 mA. (b) and (c) Show scanning SQUID microscopy images of the tricrystal region with a +1/2 and -1/2 Josephson vortex in it. $\Delta\phi_s$ is the net variation in flux through the SQUID pickup loop. 

One way to determine the total flux in a ring or vortex is to fit the SQUID microscope image to a solution, for example, of London's, Maxwell's, and the Sine-Gordon equation in the appropriate geometry \cite{kirtley1996dii,kirtley1999tdh}. Another way is to integrate the observed flux signal and multiply by a factor related to the effective area of the pickup loop, which can be calibrated, for example, assuming the Abrikosov vortices have $\Phi_0$ of flux in them. The integration is complicated by the uncertainty in assigning a value for the background field. A way to deal with this problem is to integrate over successively larger areas centered at the position of peak flux. If the flux source is localized, it is expected that a plot of integrated flux vs. integration area should saturate for large areas. The background flux can then be assigned such that this is the case. This procedure works remarkably well  \cite{tsuei2004rdp,kirtley2006arp}. Figure \ref{fig:dopesci1p}(f) Shows the integration of the total flux (in units of $\Phi_0$) of the N=+1/2 state [(b), red dots], the N=-1/2 state [(c), blue dots], and a nearby N=1 Abrikosov vortex (green dots) over a circular area A$_{\rm int}$ centered at the tricrystal point. It has been speculated that the pairing state in the cuprate superconductors could violate time reversal symmetry \cite{sigrist1995fve,sigrist1998trs,covington1997osi}, perhaps at surfaces or interfaces. The blue line in (f ) is the N=-1/2 data multiplied by -1, demonstrating double degeneracy, and time reversal symmetry for the half-flux quantum Josephson vortex. The red line in (f) is the N=1/2 data multiplied by 2, showing that the total flux at the tricrystal point is, within experimental error, half of that in a nearby Abrikosov vortex.

\begin{figure}[tb]
        \begin{center}
                \includegraphics[width=12cm]{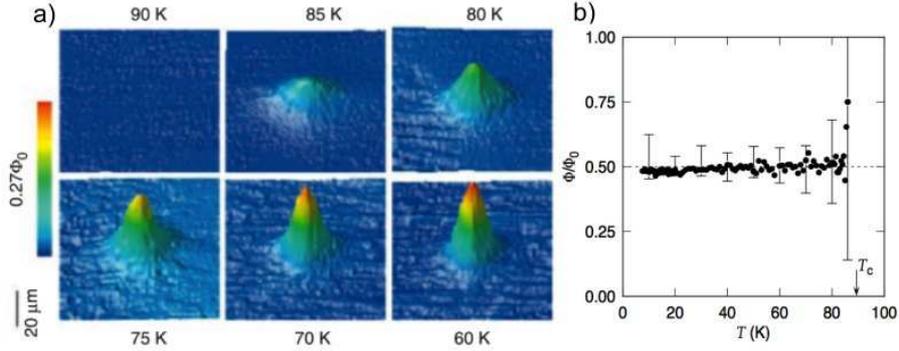}
        \end{center}
        \caption{(a) SQUID microscope images of the central point of a YBCO tricrystal sample at selected sample temperatures \cite{kirtley1999tdh} . (b) Total integrated spontaneously generated magnetic flux through the tricrystal point as a function of temperature.}
        \label{fig:temp_dependence}
\end{figure}

Imaging half-flux quantum Josephson vortices at the tricrystal point also facilitated measurements of the temperature dependence of the half-flux quantum effect. This also required the development of a SQUID microscope in which the sample could be warmed while the SQUID sensor, closely thermally coupled to the $^4$He bath, remained superconducting \cite{kirtley1999vst}.  Results from this study are shown in Figure \ref{fig:temp_dependence}. Fig. \ref{fig:temp_dependence}(a) shows SQUID microscopy images of the tricrystal point of an optimally doped YBCO sample at selected temperatures. The total integrated flux through the tricrystal point as a function of temperature is shown in Fig. \ref{fig:temp_dependence}(b). To within experimental uncertainty, the flux through the tricrystal point (Fig. \ref{fig:temp_dependence}(b)) was $\Phi_0/2$ at all temperatures between 0.5K and T$_c \sim$ 90K. 

There were several early attempts to measure the in-plane momentum dependence of the energy gap in cuprate superconductors using tunneling junctions \cite{gim1997ads,van1999esi}, but these efforts were hindered by the uncontrolled nature of the junction interfaces. Some success was achieved by Lombardi {\it et al.} \cite{lombardi2002idw}, who measured the angular dependence of the Josephson critical currents of $c$-axis tilt biepitaxial grain boundary YBCO junctions. These junctions are formed by the grain boundary between (001) and a (103) oriented films, with crystalline rotations about two axes and low interface transmission probabilities. Lombardi {\it et al.} found minima in the junction critical currents at junction angles of approximately 0$^\circ$, 35$^\circ$, and 90$^\circ$, as expected for a $d_{x^2-y^2}$ superconductor in this geometry.

The regrown YBCO ramp-edge techology \cite{smilde2002etr}, in combination with the growth of non-twinned YBCO films on vicinal SrTiO$_3$ subtrates \cite{dekkers2003myb}, allowed Smilde {\it et al.} to  produce a series of junctions with varying junction normals relative to the $a$-axis \cite{smilde2005adw}.  Junctions made with twinned YBCO films showed a 4-fold symmetry with nodes at $(2n+1)\pi/4$, $n$ an integer, expected for $d_{x^2-y^2}$ symmetry. Those made with untwinned YBCO showed nodes offset by about 5$^\circ$ from the twinned node angles, consistent with a small $s$-wave component to the gap, as expected for this orthorhombic superconductor \cite{sun1994ojp,walker1996oms,kleiner1996ptf}. Smilde {\it et al.} fit their experimental data using an in-plane gap with 83\% $d_{x^2-y^2}$, 13\% isotropic $s$-wave, and 5\% anisotropic $s$-wave pairing symmetry, resulting in a gap amplitude 50\% higher in the $b$ (Cu-O chain) direction than in the $a$-direction. 

\begin{figure}[tb]
        \begin{center}
                \includegraphics[width=14cm]{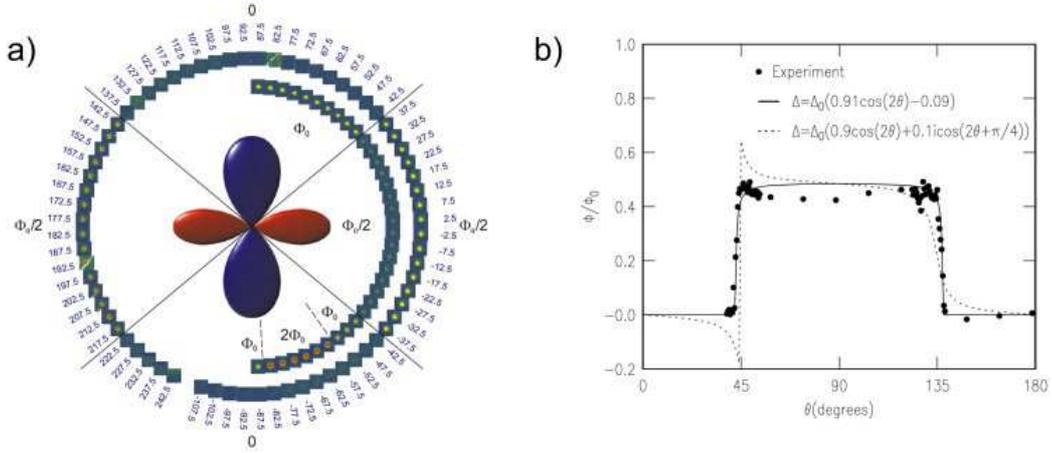}
        \end{center}
        \caption{(a) SQUID microscopy images of a series of two-junction YBCO-Nb ramp edge rings in the geometry of Fig. \ref{fig:geometries}(b).  (b) Experiment (dots) and modeling (lines) for the total integrated spontaneous flux in a set of rings, cooled in zero field, as a function of the second junction angle $\theta$. }
        \label{fig:varma}
\end{figure}

Measurements of the junction critical currents are insensitive to the orbital component of the phase of the pairing wavefunction. However, this phase component can be inferred from measurements of spontaneously generated supercurrents in YBCO-Nb rings made with the ramp-edge junction technology (see Fig. \ref{fig:geometries}(b)). Figure \ref{fig:varma} shows the SQUID microscope imaging of a series of two-junction ramp-edge YBCO-Nb rings. One of the ramp-edge angles relative to the $a$-axis direction was held fixed at -22.5$^\circ$. The second junction angle was varied in 5$^\circ$ intervals, as indicated for each image. In the outer ring of images, which correspond to a cooldown in zero field, the rings either have no or $\Phi_0/2$ spontaneous magnetization. The inner ring, corresponding to a cooldown in a finite field, shows either $\Phi_0/2$, $\Phi_0$, or 2$\Phi_0$ of flux in the rings. The transition between ring flux states occurs at angles different from $\theta_2=\pi/4+n\pi/2$ as expected for a pure $d_{x^2-y^2}$ superconductor, because of the asymmetry of the gap in the $a$ vs the $b$ in-plane directions.  Figure \ref{fig:varma}(b) shows the results for the total flux threading each ring for a second sample with an interval of 0.5$^\circ$ between values for $\theta_2$. The solid line is modeling assuming predominantly $d_{x^2-y^2}$ symmetry with a small admixture of $s$-symmetry. The $s$-wave component inferred from this fitting is smaller than that obtained by Smilde {\it et al.}'s critical current measurements, perhaps because the two-junction ring samples were slightly more twinned. The dashed line in Fig. \ref{fig:varma}(b) is theoretical modeling that includes a small imaginary component to the gap function. It is clear from this modeling that the experimental results are inconsistent with anything but a very small imaginary component to the order parameter, and therefore any time reversal symmetry breaking must also be small.

\section*{Conclusions}

Despite the considerable success of phase sensitive pairing symmetry tests in the cuprate perovskite superconductors, there have been no published phase sensitive experiments on non-cuprate unconventional superconductors except for magnetic interferometry experiments on Sr$_2$RuO$_4$ \cite{nelson2004ops,kidwingira2006dso} and the heavy fermion superconductor UPt$_3$ \cite{strand2009ecs}. In retrospect, the fact that there have been few magnetic interferometry experiments and no phase sensitive SQUID microscopy experiments reflects the fact that there were a number of favorable factors and a number of technical advances that made phase sensitive pairing symmetry experiments possible in the cuprate high-T$_c$ superconductors. As discussed above, the favorable factors included: 
1) a simple Fermi surface, 
2) a simple pairing wavefunction, 
3) grain boundaries are Josephson weak links, 
4) twin boundaries are not Josephson weak links,
5) the pairing wavefunction has odd-symmetry across twin boundaries, and
6) the Josephson coupling across several kinds of junctions favors transport normal to the interface. 

The technical advances required to make reliable phase sensitive pairing symmetry experiments possible were:
1) high quality single crystals,
2) high quality epitaxially grown films,
3) bi- tri- and quad-crystal subtrates,
4) detwinned single crystals,
5) non-twinned films on vicinal subtrates,
6) high quality ramp edge junction interfaces, and
7) high spatial resolution scanning SQUID microscopes.

Clearly more work needs to be done to apply SQUID microscopy phase sensitive pairing symmetry tests to the broad range of unconventional superconductors that are now appearing.

\section*{Acknowledgements}
I would like to thank the NanoScience Fondation for their support while writing this review.
 
\section*{References}




\bibliographystyle{elsarticle-num}
\bibliography{comren_arxiv}







\end{document}